\documentclass[epj]{svjour}

\usepackage[english]{babel}
\usepackage{verbatim}
\usepackage{amsmath}
\usepackage{amssymb}
\usepackage{graphicx}
\newcommand{\half}{\textstyle\frac{1}{2}\displaystyle}

\begin{document}

\title{Spatial Schmidt modes generated in parametric down-conversion}
\author{Filippo M. Miatto\inst{1} \and Henrique Di Lorenzo Pires\inst{2} \and  Stephen M. Barnett\inst{1} \and Martin P. van Exter\inst{2}}
\institute{SUPA and Department of Physics, University of Strathclyde, Glasgow G4 0NG, Scotland, U.K.\and Huygens Laboratory, Leiden University, PO Box 9504, 2300 RA Leiden, The Netherlands}
\date{Received: date / Revised version: date}

\abstract{
This paper presents the general spatial Schmidt decomposition of two-photon fields generated in spontaneous parametric down-conversion (SPDC). It discusses in particular the separation of the radial and azimuthal degrees of freedom, the role of projection in modal analysis, and the benefits of collinear phase mismatch. The paper is written in a review style and presents a wealth of numerical results. It aims at emphasising the physics beyond the mathematics, through discussions and graphical representations of key results. The two main conclusions of the paper are the finding of a better law to describe the effective dimensionality of the spatial part of the total Hilbert space and a possible novel feature of the radial Schmidt modes.}

\maketitle

\section{Introduction}
The understanding of processes that generate entangled states is of great importance. Current research makes a vast use of entangled states, and future technologies might rely on quantum properties of matter and fields \cite{Neilsen}. Spontaneous Parametric Down-conversion (SPDC) is one of such processes, which can generate high-dimensionally entangled states of two or more photons  \cite{Sonja:2002,Torres:2003,Molina-Terriza:2008,Franke-Arnold:2008,Miatto2012}. The literature on SPDC is well developed, however there are still some open questions, some of which we aim at answering.
Although the entanglement between SPDC photons happens also between degrees of freedom beyond the spatial ones (polarisation and time-frequency for instance \cite{Walmsley2000}), given the type of experimental setups that we refer to, we will concentrate on the spatial part of the entanglement. Therefore, the key challenge we will address in this paper is how to decompose the two-photon field $A({\bf q}_1,{\bf q}_2)$ into its Schmidt modes $u_i({\bf q}_1)$ and $v_i({\bf q_2})$:
\begin{equation}
A({\bf q}_1,{\bf q}_2) = \sum_i \sqrt{\lambda_i} u_i({\bf q}_1) v_i({\bf q_2}),
\label{schmidtgeneral}
\end{equation}
In particular, the issue of being able to take into account a phase mismatch between the photons, when calculating entanglement strength and detection probabilities has not yet found a full resolution. Approximations have been made, and with these, very interesting analytical results have been found \cite{Miatto2012,Straupe2011}. However, approximations naturally restrict the domain of validity of the results. The most common approximation used in SPDC theory (replacing a sinc function with a Gaussian function) is discussed in this paper. We present a numerical analysis of the entanglement strength, of the Schmidt modes, and of the detection amplitudes without the usual approximations. The first part of the paper is a theoretical discussion of the physical system under study and a presentation of the meaning of the Schmidt decomposition. In the second part we present the numerical results, and we describe some interesting features of the structure of the entanglement.

%%%%%%
%Theory
%

\section{Theory}
\subsection{Generated two-photon field}\label{sec:general}

We consider the generation of entangled photon pairs via the nonlinear optical process of SPDC, where single pump photons which propagate through a nonlinear crystal occasionally split in two photons of lower energy, traditionally called the signal and idler photons. We operate in the quasi-monochromatic frequency-degenerate limit, by combining a narrow-band cw pump laser at frequency $\omega_p$ with narrow-band spectral filters at the degenerate frequency $\omega_1 = \omega_2 = \omega_p/2$. When the pump field is spatially coherent and the spectral filtering is sufficiently narrow, the generated two-photon state (ignoring polarisation and therefore describing both type-I and type-II down-conversion) is pure and has the general form \cite{Mandelwolf}
\begin{equation}
| \Psi \rangle = \int \int A({\bf q}_1,{\bf q}_2) \hat{a}^\dag({\bf q}_1) \hat{a}^\dag({\bf q}_2)\,d{\bf q}_1d{\bf q}_2 | 0 \rangle ,
\label{2phot}
\end{equation}
where $\hat{a}^\dag({\bf q}_i)$ is the creation operator of a plane wave with transverse momentum ${\bf q}_i$ and $i$ is the photon label.

One of the factors of the two-photon amplitude $A({\bf q}_1,{\bf q}_2)$ in eq. \eqref{2phot}, is the phase matching function, which describes the efficiency of the down-conversion process, as a function of transverse k-vectors.
The longitudinal phase matching function in the two-photon amplitude is a sinc function, being the Fourier transform of the step function that represents the uniform amplitude of creating the photon pair along the crystal. It is usually approximated with a Gaussian function if no or very little phase mismatch (indicated in the paper by $\varphi$) is present. However, the description that derives from such approximation cannot account for an arbitrary phase mismatch. Moreover, even in the regime where $\varphi=0$, the difference in shape of the approximated amplitude from its complete form has an effect in the analysis of the down converted state, as we will show.

The the two-photon amplitude in the far field, or in momentum space, presents a ringed pattern, due to the sinc function. If the phase is matched ($\varphi=0$) the sinc gives a bright spot in the centre and secondary rings around it. Most of the intensity of the field is concentrated in the central area, it is for this reason that in case of perfect phase matching it is possible to roughly approximate the phase matching function with a gaussian function (see below for more details). In this paper we also analyse situations in which $\varphi\neq0$. In this case the argument of the sinc function is modified by a constant value, see eq. \eqref{eq:Asimple}. The effect in the amplitude for negative and decreasing $\varphi$ is to lose the central spot and obtain a more divergent field, in fact every secondary ring obtains a larger radius the lower the value of $\varphi$. On the other hand, for positive and increasing $\varphi$ the effect is to obtain an overall weaker field, less divergent, with a central spot intensity that depends on the value of $\varphi$. We will refer to the effect of tuning $\varphi$ as opening and closing the rings.

\subsubsection{Sinc phase matching}

The complete form of the two photon amplitude is \cite{Law2004}
\begin{equation} \label{eq:A}
A({\bf q}_1,{\bf q}_2) \propto {\cal E}_p({\bf q}_1+{\bf q}_2) \times {\rm sinc} \left( \half \Delta k_z L \right),
\end{equation}
where ${\cal E}_p({\bf q_1}+{\bf q_2})$ is the transverse momentum profile of the pump beam (${\bf q}_p = {\bf q}_1 + {\bf q}_2$). The function ${\rm sinc} \left( \half \Delta k_z L \right)$ quantifies the influence of phase matching, where $L$ is the crystal length and where the projected wave-vector mismatch $\Delta k_z = \Delta k_z({\bf q}_1,{\bf q}_2)$ is a function of ${\bf q}_1$ and ${\bf q}_2$.

We make the general Eq.~(\ref{eq:A}) more specific by inserting a (rotationally-symmetric) Gaussian pump profile. Other, more general, types of pump profiles can be treated analytically in the limit for a short crystal \cite{Alison2010b}. We also assume non-critical phase matching, to remove any linear dependence of $\Delta k_z$ on ${\bf q}_i$, and perform a Taylor expansion of $\Delta k_z({\bf q}_1,{\bf q}_2)$, neglecting a small term that scales with the difference between the refractive indices at the pump and SPDC wavelength multiplied by $|{\bf q}_1 + {\bf q}_2|^2$. Following the notation of Law and Eberly \cite{Law2004}, we thus rewrite Eq.~(\ref{eq:A})
\begin{equation} \label{eq:Asimple}
A({\bf q}_1,{\bf q}_2) \propto \exp{\left(-|{\bf q}_1 + {\bf q}_2|^2/\sigma^2\right)} \times {\rm sinc}\left( b^2 |{\bf q}_1 - {\bf q}_2|^2 + \varphi \right)  ,
\end{equation}
where $\sigma =2/w_p$, for a Gaussian pump profile $\mathcal E_p({\bf x}) \propto \exp{(-|{\bf x}|^2/w_p^2)}$, and
$b^2 = L/(4k_p)$, where $k_p = n \omega_p/c$ is the momentum of a pump photon in the crystal \cite{Monken1998}, and $\varphi$ is the collinear phase mismatch.

Although this challenge has been tackled mathematically by Law and Eberly \cite{Law2004}, several aspects and physical implications of the Schmidt decomposition remained undiscussed and are still a topic of active research \cite{Miatto2012,Straupe2011}. This paper aims to be as complete as possible in this discussion.

\subsubsection{Gaussian phase matching}

As anticipated, to obtain analytic solutions, many authors \cite{Miatto2012,Law2004,Monken1998} replace the sinc-type phase matching function at $\varphi = 0$ by its Gaussian approximation, using the substitution
${\rm sinc}\left( b^2 |{\bf q}_1 - {\bf q}_2|^2 \right) \approx \exp{\left( -b^2 |{\bf q}_1 - {\bf q}_2|^2\right)}$. The Gaussian approximation allows an exact Schmidt decomposition with Hermite-Gaussian or Laguerre-Gaussian eigenmodes \cite{Miatto2012,Law2004,Straupe2011,Exter2006}. The Schmidt number of this decomposition, i.e., the effective number of modes that participate in the modal decomposition, is \cite{Law2004}
\begin{equation} \label{K}
K  = \frac{1}{4}\left(b\sigma+\frac{1}{b\sigma}\right)^2 .
\end{equation}
Most experiments operate in the weak-focusing limit (large $w_p$, long Rayleigh range), where $b \sigma \ll 1$ and $K \approx 1/(2 b \sigma)^2$. The generic beam waist of all Hermite-Gauss (HG) or Laguerre-Gauss (LG) modes, defined by the fundamental mode $u_{0,0}({\bf q}) = \exp{(-q^2/q_0^2)}$, is $q_0 = \sqrt{\sigma/(2b)}$ being the geometric mean between the width of the pump profile and the Gaussian phase matching function.

The eigenmodes of the Gaussian Schmidt decomposition are not unique (because of the degeneracy of all the eigenmodes belonging to a fixed mode order) and, for example, can equally well be taken as the Hermite-Gaussian or Laguerre-Gaussian set of modes. 

\begin{figure}[ht]
\resizebox{0.95\columnwidth}{!}{%
  \includegraphics{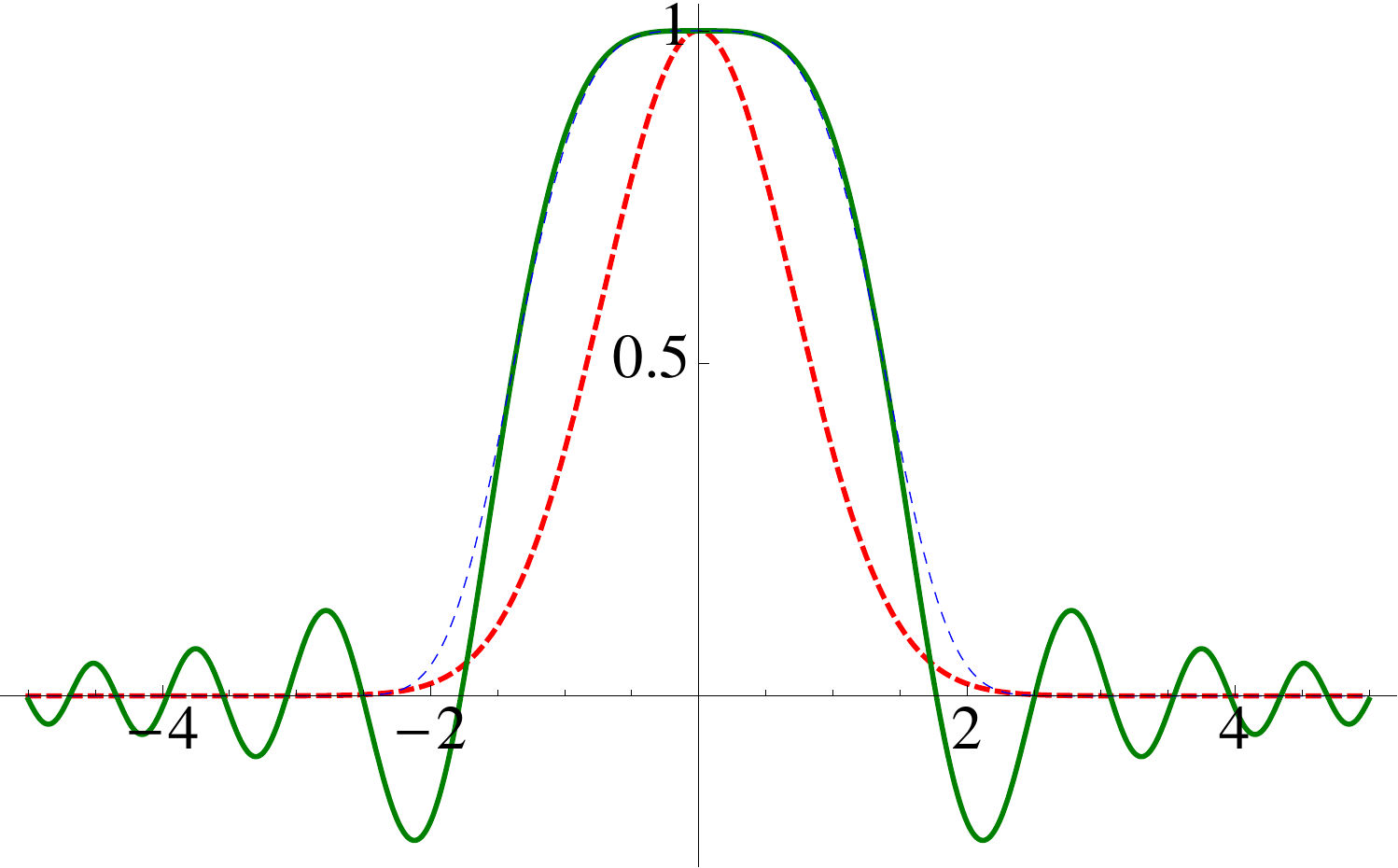}
}

\caption{\label{sincgauss}(Color online) Comparison between the three different phase matching functions. The solid line is the full sinc form $\mathrm{sinc}(q^2)$, the dashed lines are the gaussian (red, narrower) $\exp(-q^2)$ and optimised supergaussian (blue) approximation $\exp(-0.193q^4)$. The similarity between the Gaussian and sinc functions is questionable, as it is the similarity between supergaussan and sinc for the overall scaling and for the oscillations at  $q\gtrsim2$.}
\end{figure}

The numerical results presented in section 3 suggest that the Gaussian expansion can be improved somewhat by writing ${\rm sinc}\left( b^2 |{\bf q}_1 - {\bf q}_2|^2 \right) \approx \exp{\left( -\alpha^2 b^2 |{\bf q}_1 - {\bf q}_2|^2\right)}$, where $\alpha$ is a (scaling) constant chosen such that both functions satisfy some common criterium. This modified scaling results in a replacement of $b \rightarrow \alpha b$ in all  expressions that originate from the Gaussian approximation. We postpone the discussion on the rescaling factor $\alpha$ at the end of the paper, as additional concepts have to be introduced. In Figure \ref{sincgauss} we plot the sinc phase matching (solid green line), comparing it to its Gaussian  (red dashed) and supergaussian (blue dashed) approximations. It is straightforward to notice that the Gaussian approximation $\exp(-q^2)$ scales very rapidly, thus missing out high-$q$ modes. The sinc function $\mathrm{sinc}(q^2)$, instead, has a $1/q^2$ overall scaling, letting high-$q$ modes have an effect on the Schmidt decomposition. The supergaussian $\exp(-bq^4)$ with $b=0.193$ is a better approximation than the Gaussian, as it captures the central scaling of the sinc much better \cite{Walmsley2003,Walmsley2010}. This is due to the Taylor expansion of the supergaussian having the same powers as the sinc. However, there are two main problems associated with the supergaussian approximation. The first is that although it models the central scaling for a wider range of $q$ values, it misses the $1/q^2$ overall scaling too, failing to carry high-$q$ modes, and it does not display any ring structure. The second problem is that analytically it is not an advantage: for perfect phase matching it is possible to find angular Schmidt modes in terms of Bessel functions of two variables, which are defined in \cite{korsch,dattoli}, but as the literature on generalised Bessel functions is poorer than the literature on single variable Bessel functions, it is perhaps an unprofitable effort to try to find the full analytical Schmidt decomposition using the supergaussian approximation, and as we will show in section 3.3, a heuristic rescaling of the Gaussian approximation is sufficient to express the Schmidt number correctly in case of perfect phase matching.

\subsection{Schmidt decomposition and Entanglement}
For the arguments in this paper it is sufficient to recall that the Schmidt decomposition of an entangled state is its representation in the joint basis that maximises the mutual information. Notice that in Eq. (\ref{schmidtgeneral}) the joint modes are the product of one mode per subsystem (hence each Schmidt mode is separable). The decomposition, in fact, also specifies the weights $\sqrt{\lambda_i}$ of single joint modes $u_i(\mathbf{q}_1) v_i(\mathbf{q}_2)$, the square of which is interpreted as the normalised probability (via $\sum_\ell \lambda_\ell=1$) of detecting the $i$-th joint mode, were we to measure in the Schmidt basis. The more the distribution of the probabilities $\{\lambda_i\}$ is ``spread out'' the more entangled the state is. This leads to the natural choice of choosing entropic measures to quantify entanglement. In our case we choose the Schmidt number $K=1/\sum_i \lambda^2_i$, where the probabilities are normalised: $\sum_i \lambda_i =1$.

The Schmidt decomposition of the state (\ref{2phot}) can be thought as made of two stages, because we are working with two pairs of coordinates: an angular coordinate per subsystem and a radial coordinate per subsystem. In the first stage we separate the angular variables, in the second stage we separate the radial variables. The decomposition will have therefore two sums, and the weights and joint modes will have two indices \cite{Miatto2012}.

We start with the two-photon amplitude $A(\mathbf{q}_1,\mathbf{q}_2)$ and we write it in polar coordinates (which we call $q_{1,2}$ and $\theta_{1,2}$). The rotational symmetry of $A(\mathbf{q}_1,\mathbf{q}_2)$ limits the angular dependance to a function of $(\theta_1-\theta_2)$ only.
The orthogonality of the azimuthal functions $e^{i \ell (\theta_1 - \theta_2)}$ allows for an easy extraction of the quantity
\begin{equation} \label{eq:Fl}
\sqrt{P_\ell} F_\ell(q_1,q_2) =\int_0^{2\pi} A({\bf q}_1,{\bf q}_2)e^{-i \ell (\theta_1 - \theta_2)} d (\theta_1 - \theta_2) .
\end{equation}

This means that with a Fourier transform we can write the two photon amplitude in a joint basis of the Orbital Angular Momentum (OAM):
\begin{equation}
A(\mathbf{q}_1,\mathbf{q}_2)= \frac{1}{2\pi}\sum_\ell \sqrt {P_\ell} F_\ell(q_1,q_2)e^{i\ell\theta_1}e^{-i\ell\theta_2}
\label{angulardecomp}
\end{equation}
Where $P_\ell$ is the probability of measuring a pair of photons with OAM $\ell$ and $-\ell$, were we to measure in the joint OAM basis.
The radial amplitude is normalised as $\int|F_\ell(q_1,q_2)|^2q_1q_2\,dq_1dq_2=1$.
In this way we have separated the two angular variables, and each joint OAM mode is labelled by an integer number $\ell$.
We can therefore proceed to the second stage and consider the functions $\sqrt{P_\ell}F_\ell(q_1,q_2)$ and separate the radial variables. In order to make this separation easy and truly one-dimensional, we follow the reasoning of Law and Eberly \cite{Law2004} by introducing a prefactor $\sqrt{q_1 q_2}$, to account for the Jacobian of the transformation $\vec{q} \rightarrow (q,\theta)$, and separating the radial function as
\begin{equation}
\sqrt{q_1 q_2} \sqrt{P_\ell} F_\ell(q_1,q_2) = \sum_{p \geq 0} \sqrt{\lambda_{\ell,p}} \phi_{\ell,p}(q_1) \phi_{-\ell,p}(q_2) .
\label{radialdecomp}
\end{equation}
Symmetry considerations also lead to the separated functions to be equal, i.e. we use $\phi(q_1)$ as well as $\phi(q_2)$.
The prefactor $\sqrt{q_1 q_2}$ is needed to facilitate the simple normalisation of the \hyphenation{ei-gen-fun-ctions} eigenfunctions ($\int \phi_{\ell,p}(q) \phi_{l,p'}(q) dq = \delta_{p,p'}$) that is required for the numerical decomposition that we discuss later on. For convenience, we will also define $u_{\ell,p}(q) \equiv \phi_{\ell,p}(q) / \sqrt{q}$ such that $\int u_{\ell,p}(q) u_{l,p'}(q) q dq = \delta_{p,p'}$.

Note that we had to introduce the non negative, integer label $p$. Such quantum number labels radial eigenmodes. The numerical analysis will lead to some more insight on this quantum number, as we will show in section~\ref{numerics}.
After putting these two results together we obtain the full two-dimensional Schmidt form of the two-photon amplitude
\begin{equation} \label{eq:ASchmidt}
A({\bf q}_1,{\bf q}_2) = \sum_{\ell=-\infty}^{\infty} \sum_{p=0}^\infty\sqrt{\lambda_{\ell,p}} u_{\ell,p}({\bf q}_1) u_{-\ell,p}({\bf q}_2) .
\end{equation}
Note that this form is analogous to Eq. (\ref{schmidtgeneral}) with $u_{i}(\mathbf{q})\equiv u_{\ell,p}(q)e^{i\ell\theta}$.

The radial shape $\phi_{\ell,p}(q)$ of these profiles depends \hyphenation{stron-gly} strongly on the collinear phase mismatch $\varphi$; it resembles the well-known Laguerre-Gaussian ($\ell,p$)-profiles only for $\varphi \approx 0$, but deviates strongly from these profiles at $\varphi \neq 0$ (see below). As anticipated, the effective number of spatial modes (which can be thought of as the strength of the entanglement) is given by the Schmidt number
\begin{equation}
K = 1/\sum_{\ell,p} \lambda_{\ell,p}^2
\label{fullk}
\end{equation}
where the probabilities are normalised: $\sum_{\ell,p} \lambda_{\ell,p}= 1$.

A simple scaling argument shows that the outcome of the decomposition depends only on two dimensionless parameters. The first parameter is the product $b\sigma = \sqrt{\frac{L}{w_p^2k_p}}=\sqrt{L_R/2}$ where we define $L_R$ to be the crystal thickness normalised to the Rayleigh range of the pump beam, $L_R\equiv L/z_R$. The second parameter is the collinear phase mismatch $\varphi$.

\subsubsection{Entanglement in orbital angular momentum}

The rotational symmetry of the considered geometry and the associated conservation of OAM makes the azimuthal component of the entanglement easier to address and more fundamental than its radial counterpart. Hence, most experiments have concentrated on the OAM part of the spatial entanglement. In this context we are interested in the probability $P_\ell$ (which appears in Eq. (\ref{angulardecomp})) of detecting a pair of photons with OAM, respectively, $\ell$ and $-\ell$. Such probability is related to the weights $\lambda_{\ell,p}$ via $P_\ell = \sum_p \lambda_{\ell,p}$.  $P_\ell$ is called the \textit{spiral weight}, its full distribution $\{ P_\ell\}$ is called the \textit{spiral spectrum}, and the width of this distribution is called the \textit{spiral bandwidth} \cite{Torres:2003}. The dimensionality of the generated OAM entanglement is defined by the azimuthal Schmidt number
\begin{equation}
K_{\mathrm{az}} = 1/\sum_\ell P_\ell^2 .
\end{equation}
To be compared to the full Schmidt number (\ref{fullk}).
In case of high-dimensional spatial entanglement ($K \gg 1$) and $\varphi \approx 0$, the relation between both forms of entanglement is $K_{\rm az} \approx 2 \sqrt{K}$, as the effective range of the $\ell$ labels is typically $\approx 4 \times$ as large as that of the $p$ labels. This statement is exact for the Gaussian approximation of the two-photon field, where \cite{Exter2006}
\begin{equation} \label{eq:Plp}
\lambda_{\ell,p} = C_1 \exp{\left( -(2p + |\ell|)/C_2\right)} ,
\end{equation}
where $C_1$ and $C_2$ are two constants. It does, however, not apply to the full expression for $\varphi\neq0$. 
Our numerical approach allows us to analyse what happens also in such regime.

\subsection{Optical etendue}\label{sec:etendue}

The Schmidt decomposition of the two-photon field is equivalent to a coherent mode decomposition of the coherence function of the reduced one-photon state. The reduced one-photon operator $\hat\rho^{(1)}$, obtained by tracing $| \Psi \rangle \langle \Psi |$ over all possible states of the second photon, is an incoherent (and weighted) mixture of all of all Schmidt 
states. The extent to which the reduced density operator is mixed reflects the degree of entanglement for the pure two-photon state $|\psi\rangle$. The Schmidt number $K$ can thus be estimated by comparing the one-photon coherence of the SPDC emission with that of more standard optical sources.

At sufficiently large $K \gg 1$, the reduced one-photon field is quasi-homogeneous, i.e., the reduced one-photon density matrix $\rho^{(1)}({\bf q}_1,{\bf q}_1')$ factorizes in a product of an intensity function of sum coordinate $({\bf q}_1 + {\bf q}_1')$ and a coherence function of difference coordinate $({\bf q}_1 - {\bf q}_1')$. The Schmidt number $K$ of the quantum state then reduces to the normalised optical etendue $N \equiv A \Omega/\lambda^2$ of the one-photon field, where $A$ is the effective area of the source and $\Omega$ is its effective (solid) opening angle \cite{Exter2006,Pires2009c}. The optical etendue quantifies the effective number of transverse modes contained in a partially-coherent beam. The relation $K \approx N$ becomes exact if the source is \hyphenation{qua-si-ho-mo-ge-ne-ous} quasi-homogeneous and if the effective area $A$ and opening angle $\Omega$ are defined in a convenient way \cite{Pires2009c}.

The relation between $K$ and $N$ yields an easy and intuitive interpretation of the Schmidt dimension. For this, we convert the opening angle of the source $\Omega$ into a transverse coherence width $w_{\mathrm{coh}}$ in the source plane, which we conveniently define as
\begin{equation}
w_{\mathrm{coh}} \equiv 4 b = \sqrt{\frac{L \lambda_0}{\pi n}} ,
\end{equation}
where $L$ is the crystal length, $\lambda_0$ is the emission wavelength in vacuum and $n$ is the refractive index. The transverse coherence width $w_{\mathrm{coh}}$ of the one-photon field is similar to the width of the two-photon coherence function $V(\vec{x}_1 - \vec{x}_2)$ defined in refs.~\cite{Pires2009b,Pires2009c} as the Fourier-transform of the sinc-type phase matching function. With the above definition of $w_{\mathrm{coh}}$, the Law-and-Eberly expression for the Schmidt number at $b \sigma \ll 1$ reduces to the logical form
\begin{equation}\label{eq:Kcoh}
K \approx \left( \frac{1}{2 b \sigma} \right)^2 = \left( \frac{w_p}{w_{\mathrm{coh}}} \right)^2 ,
\end{equation}
which shows that the Schmidt number $K$ simply counts the number of ``independent" coherent regions in the source. The above definition of $w_{\mathrm{coh}}$ also provides for an easy rewrite of the results of the Gaussian expansion discussed in subsection (2.1.2). The width $w_0$ of the fundamental Gaussian Schmidt mode $u_{0,0}(\vec{q}) \propto \exp{-|\vec{q}|^2/q_0^2}$ in real space is $w_0 = 2/q_0 = \sqrt{w_p w_{\mathrm{coh}}}$, being the geometric mean between the pump waist and the transverse coherence length of the generated field. The modal amplitude of the Gaussian $u_{\ell,p}$ mode is 
\begin{equation}
\sqrt{\lambda_{\ell,p}} \approx \exp{[-(2p+|\ell|)/K_{1D}]} ,
\end{equation}
where $K_{1D} = w_p/w_{\mathrm{coh}}$ is the one-dimensional equivalent of the Schmidt number.

\subsection{Projective measurement of OAM entanglement}\label{sec:OAM}

In this subsection, we will briefly compare two experimental techniques that have been developed to characterise in particular the OAM contents of two-photon sources. We distinguish between measurements with bucket detectors, which record the complete field, and measurements with single-mode detectors, which record projected components of the field. 

A Schmidt analysis of the two-photon field with bucket detectors, which by definition have no spatial sensitivity, requires two-photon interference before measurement. More specifically, one should measure the visibility of the famous two-photon (Hong-Ou-Mandel) dip \cite{HOM} as a function of the relative alignment of the interfering beams. Measurements with one beam rotated with respect to the other provide complete and unbiased information on the azimuthal part of the entanglement \cite{Zambrini2006,Pires2010}, as the visibility of the observed two-photon interference is 
\begin{equation}\label{eq:V}
V(\Delta {\tilde \theta}) \propto \int \int A^*(q_1,q_2;\Delta \theta) A(q_1,q_2;\Delta \theta + \Delta {\tilde \theta}) q_1 q_2 dq_1 dq_2 \nonumber
\end{equation}
\begin{equation}
= \sum_{\ell=-\infty}^{\infty} P_\ell \exp{(i\ell\Delta {\tilde \theta})},
\end{equation}
where ${\tilde \theta}$ is the rotation angle between the interfering two-photon fields, which is twice the rotation angle applied in a single arm of the interferometer. This angular dependence originates from the interference between two-photon fields with either the $\ell$ or the $-\ell$ photon in arm 1 and the other photon in arm 2. Equation (\ref{eq:V}) is based on the entanglement between these contributions and the natural symmetry $P_{-\ell} = P_\ell$.

A Schmidt analysis of the two-photon field with single-mode detectors, each comprising a single-mode fiber positioned in front of a detector, requires some form of mode transformation before projection. This mode transformation is typically performed with (a set of) fixed phase plates or an (adjustable) spatial light modulator (SLM). The combined transformation-projection technique has been used successfully for the analysis of the OAM contents of the two-photon field \cite{Pires2009c}. The generated two-photon field is then typically projected onto two modes of the form $\phi_d(q) \exp{(i f(\theta))}$, where $\phi_d(q)$ is the (image of the) mode profile of the fiber and $\exp{(i f(\theta))}$ is the phase profile imposed by the phase-transforming element, which typically has the standard form $\exp{(i\ell\theta)}$. The rotational symmetry of the generated field $A({\bf q}_1,{\bf q}_2)$, which manifests itself in the generation of photon pairs with opposite OAM only, allows one to derive simple expressions for the projected two-photon field. It is easy to show that the maximum information contained in these projections, for any combination of azimuthal phase profiles $f(\theta)$, is contained in the function
\begin{align}
g_{\rm proj}(\Delta \theta) &= \int \int A({\bf q}_1,{\bf q}_2) u_d^*(q_1) u_d^*(q_2) q_1 q_2 dq_1 dq_2 \nonumber\\
&\propto \sum_{\ell=-\infty}^{\infty} \sum_{p=0}^\infty C_{\ell,p} \sqrt{\lambda_{\ell,p}} \exp{(i\ell\Delta \theta)},
\end{align}
where $\Delta \theta \equiv \theta_1-\theta_2$. The projection coefficients
\begin{equation}
C_{\ell,p} = | \int u_{\ell,p}(q) u_d(q) q \,dq|^2 \leq 1 ,
\end{equation}
quantify the spatial overlap between each generated Schmidt mode and detector mode. They also quantify the bias imposed by the mode projections. 

As an example, let us consider the projection on a detector mode $u_d(q)$ that is much wider than a range of lower-order $u_{\ell,p}(q)$ modes. Under these conditions, the projection coefficients will have a strong bias towards the $p=0$ modes, as all $u_{\ell,p}(q)$ with $p \geq 1$ modes exhibit at least one oscillation (in the radial direction) making $C_{\ell,p \neq 0} \ll C_{\ell,p=0}$. Hence, projection experiments are mainly sensitive to the $p = 0$ components of the two-photon field. A calculation of the projection-imposed bias on the $\ell$ modes is more complicated. A rigorous analysis, in terms of Lerch transcendent functions, is presented in another contribution to this issue \cite{Miatto2012b}.% We can, however, already make a very rough approximation of the dependence of the projection coefficients $C_{\ell,0}$ with $\ell$. Under the mentioned condition of a ``very wide detection mode", we expect $C_{\ell,0}$ to increase with $\ell$ at modest $\ell$ values, where the modal overlap between the Laguerre-Gaussian modes and the wide detection mode $u_d(q) \exp{(i \ell \theta)}$ increased approximately as $C_{\ell,0} \propto \sqrt{|\ell|+1}$. We expect $C_{\ell,0}$ to decrease at large $\ell$, where the Schmidt modes become too large for the detection mode. This prediction is quite speculative and needs to be checked by a rigorous analysis. It does, however, indicate that the OAM decomposition measured via mode projection is strongly biased and rather tends to increase the measured azimuthal Schmidt number $K_{\mathrm{az}}$ than to decrease it.

%%%%%%%%%%
%Numerical results
%

\section{Numerical results}\label{numerics}

In this section we will presents numerical results for the general Schmidt decomposition of Eq.~(\ref{eq:Asimple}) into Eq.~(\ref{eq:ASchmidt}). We will in particular discuss the distribution of the modal weights $\lambda_{\ell,p}$, and their separation over the azimuthal and radial degrees of freedom, the shape of the Schmidt modes $u_{\ell,p}(q)$, and the influence of the collinear phase mismatch $\varphi$.

The numerical analysis is relatively straightforward. We start from the generated two-photon field $A({\bf q}_1, {\bf q}_2)$ defined by Eq.~(\ref{eq:Asimple}) and perform an OAM decomposition into radial-only function $\sqrt{P_\ell}F_\ell(q_1,q_2)$ using Eq.~(\ref{eq:Fl}). The analytical integration is performed using an approximation that works well for $b\sigma\lesssim0.2$, which is the most relevant experimental situation as it corresponds to a weakly focused pump. Employing the analytical integration of the angular variable yields analogous final results in much less time, as the numerical methods consist only in a diagonalisation of matrices. The prescribed radial integration is performed on a discrete equidistant grid of $q$-values. After multiplication by a factor $\sqrt{q_1 q_2}$, to account for the Jacobian of the transformation, we obtain the scaled $N \times N$ matrix $F$ with a dimension set by the required accuracy ($N$ is typically in the order of the hundreds). The Schmidt decomposition in Eq. (\ref{radialdecomp}) of $\sqrt{P_\ell}F_\ell(q_1,q_2)$ is equivalent to the diagonalisation of the matrix $F$.
To avoid potential complications associated with the two-particle nature of the
$\phi(q_1)\phi(q_2)$ phase space, we first multiply $F$ by its transpose $F^\dag$,
in order to obtain the equivalent of the reduced one-photon density matrix, 
and diagonalize the matrix $FF^\dag = F^2$ instead. The diagonalisation of the matrix $F^2$ yields three matrices: $F^2=MEM^T$, two of which represent the modes ($M$) and one of which is diagonal $(E)$ and represents the eigenvalues. The resulting eigenvalues 
are $\lambda_{\ell,p}$ (at fixed $\ell$); the resulting eigenmodes are $\phi_{\ell,p} =
\sqrt{q} u_{\ell,p}(q)$.

In the following subsection we present the weights in the matrix $E$ and we show what are the consequences of having a nonzero phase mismatch.
Similarly, we also present the modes in the matrix $M$, and we show the consequences of having a nonzero phase mismatch.

\subsection{Schmidt weights}

As a typical example we will concentrate on the case $b\sigma = 0.05$, which for instance corresponds to a 2~mm thick PPKTP pumped by a $w_p =230~\mu$m diameter pump laser at a pump wavelength of 413~nm.
\subsubsection{Phase mismatch $\varphi=0$}

\begin{figure}[ht]
\resizebox{0.95\columnwidth}{!}{%
  \includegraphics{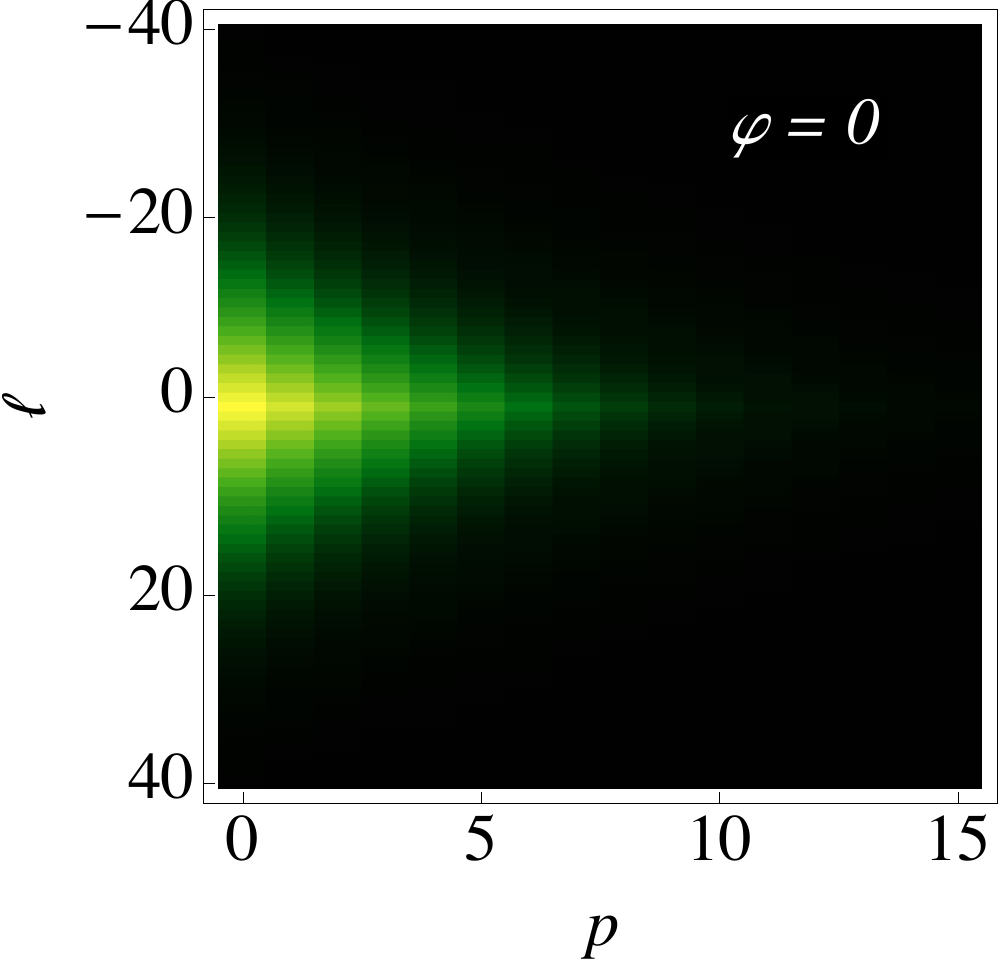}
}
\caption{False color plot of the modal spectrum $\lambda_{\ell,p}$ (with $p$ horizontal and $\ell$ vertical) for $b\sigma=0.05$ and $\varphi = 0$. Note the typical triangular form and the $\approx 4 \times$ larger range of relevant $\ell$  values as compared to $p$-values.}
\label{fig:spectrum}
\end{figure}

Figure \ref{fig:spectrum} is a false-color plot of the probabilities $\lambda_{\ell,p}$ of the modal decomposition for $b\sigma = 0.05$ and $\varphi = 0$. The Schmidt numbers deduced from this calculation are $K \approx 231$ and $K_{\mathrm{az}} \approx 32$, respectively. These numbers satisfy the approximate relation $K_{\mathrm{az}} \approx 2 \sqrt{K}$ associated with the triangular mode spectrum in figure~\ref{fig:spectrum}. They are, however, considerably larger than the values of $K \approx 100$ and $K_{\mathrm{az}} \approx 20$ expected from a simple Gaussian expansion based on the approximation ${\rm sinc}(q^2) \approx \exp{(-q^2)}$. The reason for this difference, as anticipated, is the fact that the Gaussian approximation is missing high-$q$ transverse modes.

\subsubsection{Phase mismatch $\varphi<0$}
We consider now geometries with non-perfect phase matching and in particular the case $\varphi < 0$ for which the SPDC rings have opened up.
\begin{figure}[ht]
\resizebox{0.95\columnwidth}{!}{%
  \includegraphics{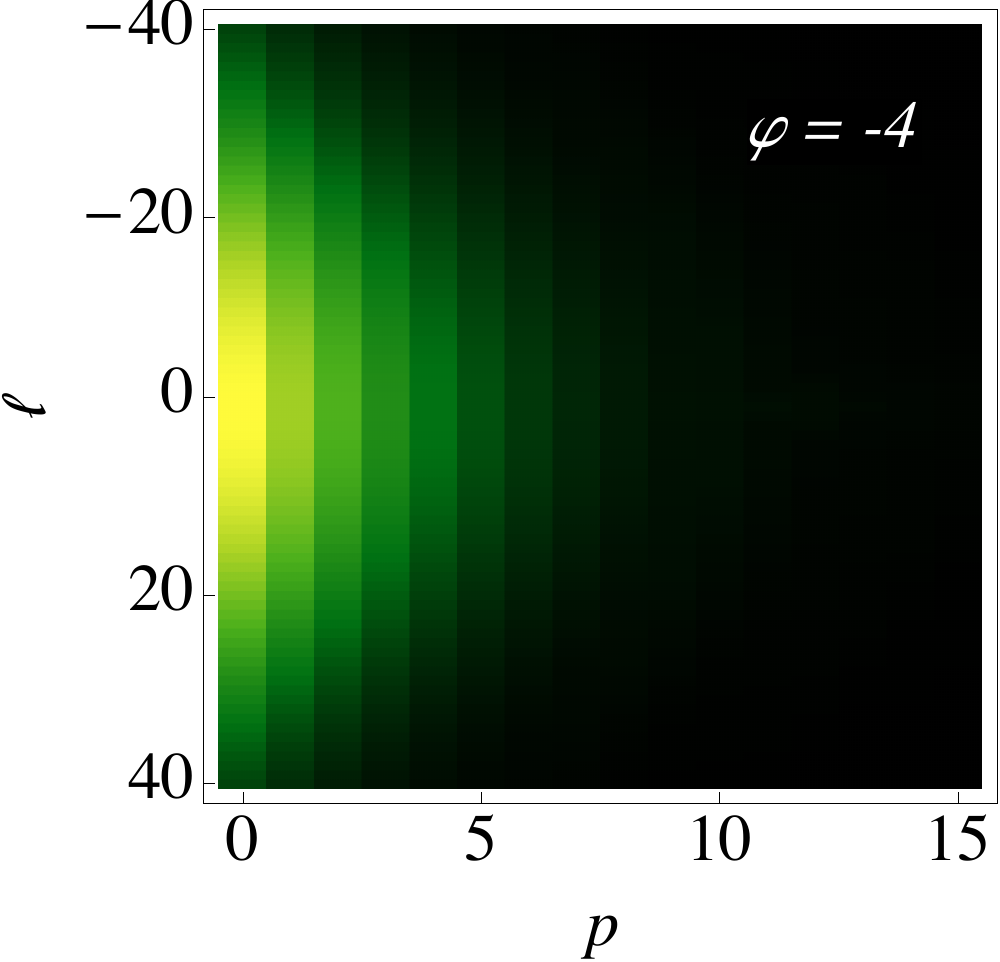}
}
\caption{False color plot of modal spectrum $\lambda_{\ell,p}$ (with $p$ horizontal and $\ell$ vertical) for $b\sigma=0.05$ and $\varphi = -4$. Note the strongly elongated form in the $\ell$ direction and the reduced width in $p$.}
\label{fig:spectrum_detuned}
\end{figure}

Figure \ref{fig:spectrum_detuned} is a false colour plot of the probability $\lambda_{\ell,p}$ of the modal decomposition calculated for $b\sigma = 0.05$ and $\varphi = -4$. On account of phase mismatch, the full 2-dimensional Schmidt number has increased by a factor $\approx 2 \times$ to $K \approx 425$. This increase coincides with a similar increase in the space angle and the angular integrated output of the SPDC source. Most noticeably, the spread in OAM values is now much larger then in the $\varphi =0$ case, at the expense of the spread in $p$ values. The azimuthal Schmidt number has increased from its prior value of 32 at $\varphi=0$ to $K_{\rm az} \approx 72$. 

\subsubsection{Phase mismatch $\varphi>0$}
For the sake of completeness we consider now the case of positive phase mismatch, $\varphi > 0$. In this case the SPDC rings have closed, and the efficiency of the down-conversion process is considerably lower, normally by two orders of magnitude, which degrades the S/N ratio by a large extent.
\begin{figure}[ht]
\resizebox{0.95\columnwidth}{!}{%
  \includegraphics{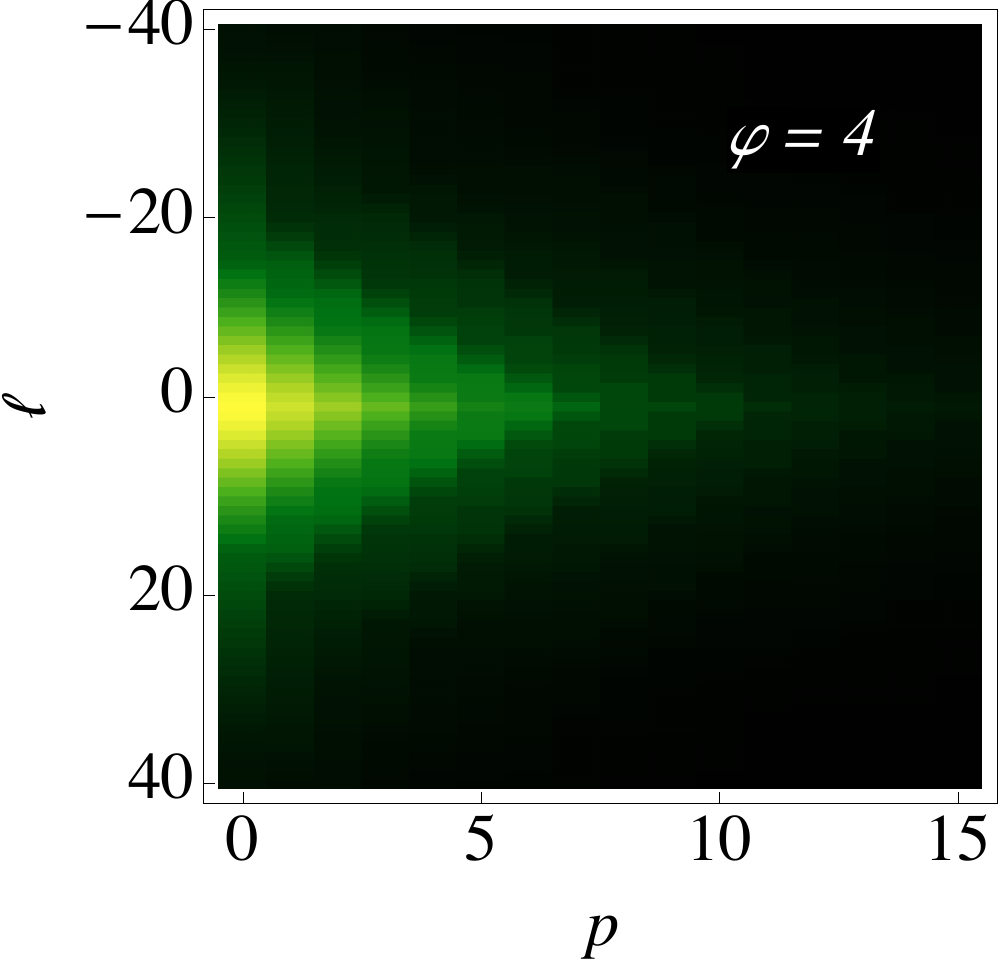}
}
\caption{\label{fig:spectrum_detuned2}False color plot of modal spectrum $\lambda_{\ell,p}$ (with $p$ horizontal and $\ell$ vertical) for $b\sigma=0.05$ and $\varphi = 4$. Note the strongly elongated form in the $p$ direction and the reduced width in $\ell$ as compared to the case $\varphi = - 4$.}
\end{figure}

Figure \ref{fig:spectrum_detuned2} is a false color plot of the eigenvalue $\lambda_{\ell,p}$ of the modal decomposition calculated for $b\sigma = 0.05$ and $\varphi = 4$. As compared to the case of negative detuning, we find that the width in the $p$ direction has increased at the expense of the width in the $\ell$-direction. In this example we calculated values of $K\approx625$ and $K_{az}\approx57$.
 
The optical etendue argument presented in Sec.~(\ref{sec:etendue}) allows one to explain why the spatial entanglement gets concentrated in the azimuthal degree of freedom, instead of the radial one, for $\varphi < 0$ and why this concentration hardly occurs for $\varphi > 0$. Our geometric argument is based on the notion that the transverse coherence width in the azimuthal and radial directions are comparable, being Fourier related to the size of the pump beam. At negative detuning $\varphi < 0$, where the fundamental SPDC ring opens up, the circumference of this SPDC increases with phase mismatch while the radial thickness of the ring decrease accordingly. As a result, the number of spatial modes that fit in the azimuthal ($\ell$) direction, increase while the number of modes in the radial ($p$) direction decrease. The total area of the open SPDC ring at $\varphi \ll -1$ is approximately twice as large as the area of central emission at $\varphi =0$, making the 2-dimensional $K$ also about two times larger. At positive detuning $\varphi > 0$ the fundamental SPDC ring disappears and only weak secondary rings remain. As these rings are numerous and have similar intensities, the number of spatial modes in the Schmidt decomposition can be quite large and is more evenly distributed over $\ell$ and $p$ mode numbers.

\begin{figure}[ht]
\resizebox{\columnwidth}{!}{%
  \includegraphics{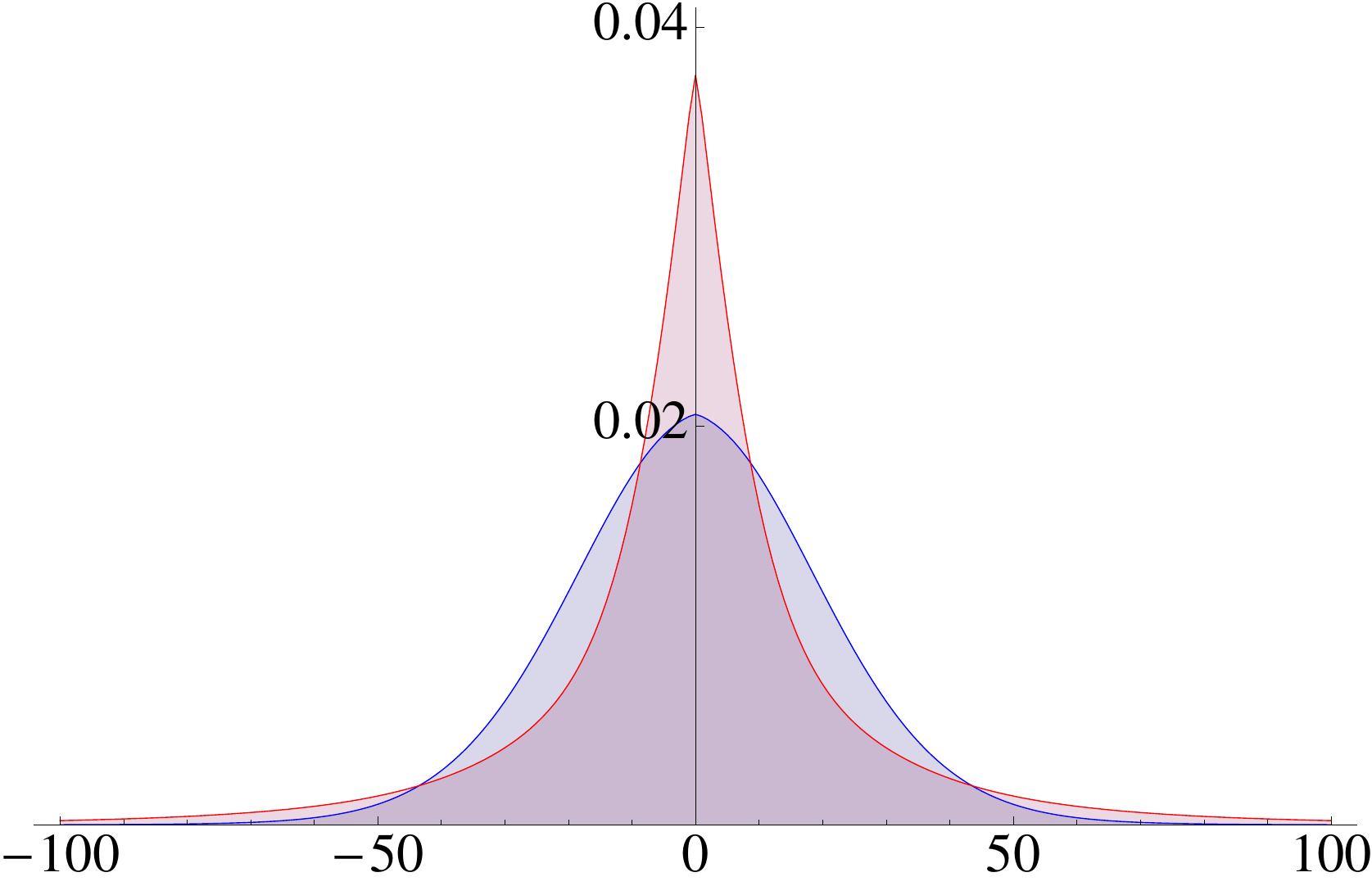}
}
\caption{\label{Kaz} Probability distributions of OAM modes (i.e. summing over $p$). The Lorenzian-like distribution (red) is for positive mismatch and the Gaussian-like distribution (blue) is for negative phase mismatch.}
\end{figure}

Next we single out the OAM part of the entanglement for the considered cases $\varphi = 0, -4$ and $4$. We do so by summing the modal weights $\lambda_{\ell,p}$, depicted in figures \ref{fig:spectrum}, \ref{fig:spectrum_detuned} and \ref{fig:spectrum_detuned2}, over its radial quantum number $p$. What one then obtains are three different probability distributions that show different behaviour over the same range of $\ell$ values. Figure \ref{Kaz} shows the modal weights ($\sum_p \lambda_{\ell,p}$) of the OAM modes. Note that the behaviour changes from Lorenzian-like to Gaussian-like as the phase mismatch goes from positive to negative values \cite{Pires2010}. We stress that the case of positive detuning is experimentally unfavorable on account of the limited brightness of a source where the fundamental SPDC ring has closed/disappeared.

\subsection{Schmidt modes}
In this subsection we will present a series of numerical results that show the radial profile of the Schmidt modes in momentum space. The profile of the modes is influenced by the presence of a phase mismatch, and as anticipated the modes resemble the LG modes only at $\varphi=0$.

\subsubsection{Phase mismatch $\varphi=0$}
\begin{figure}[ht]
\resizebox{\columnwidth}{!}{%
  \includegraphics{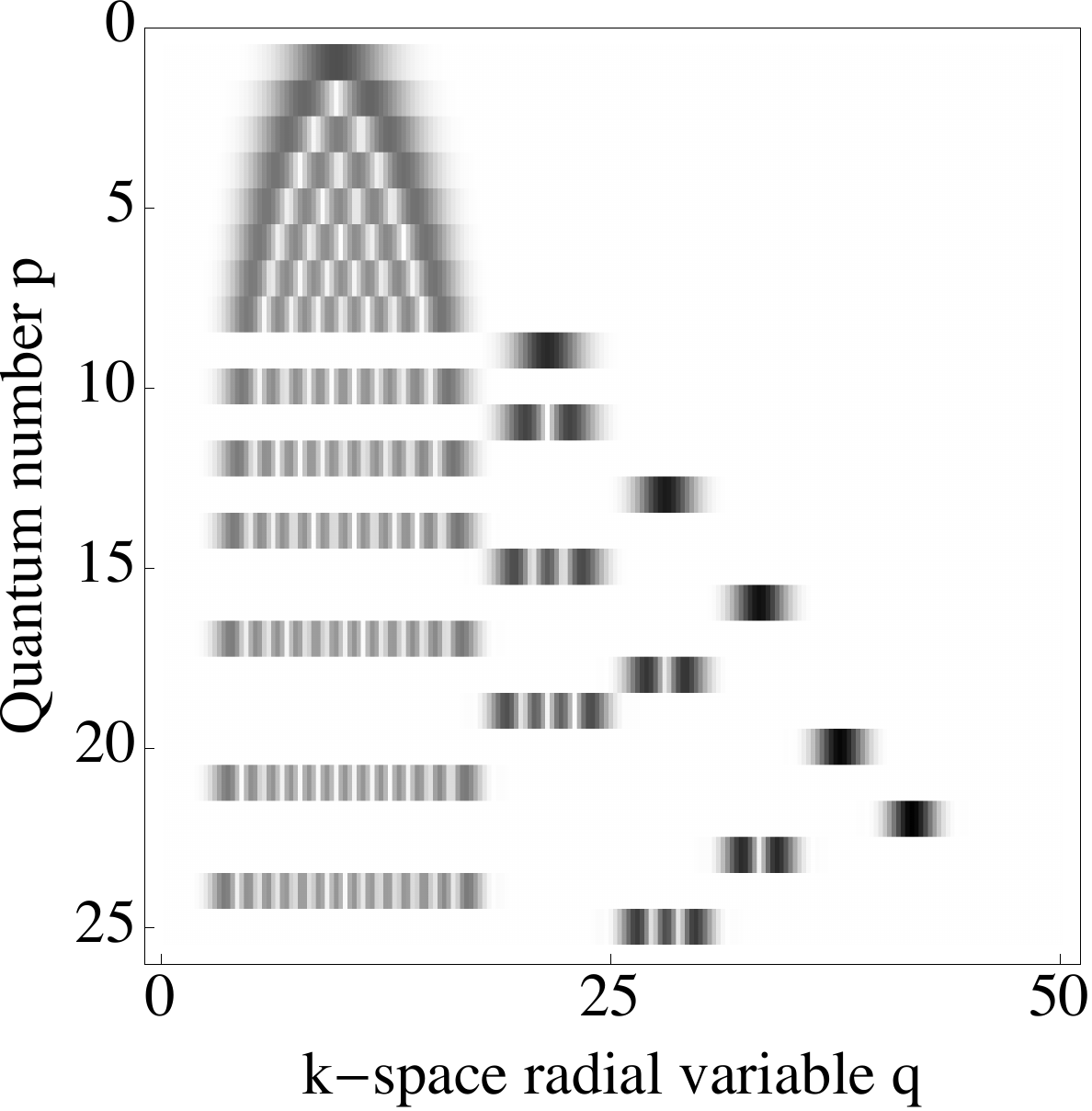}
}
\caption{\label{fig:eigenmodes}A part of the matrix $M$ for $\ell=10$ is plotted in a black and white array plot. The intensity profile $|\phi_{l,p}(q)|^2$ of a single radial eigenmode is represented in each horizontal series. The radial quantum number $p$ labels the different series. They are organised from top to bottom in order of contribution to the total wave function. In this example the radial field that was decomposed was the one in figure \ref{radialfield1}.}
\end{figure}

\begin{figure}[ht]
\resizebox{\columnwidth}{!}{%
  \includegraphics{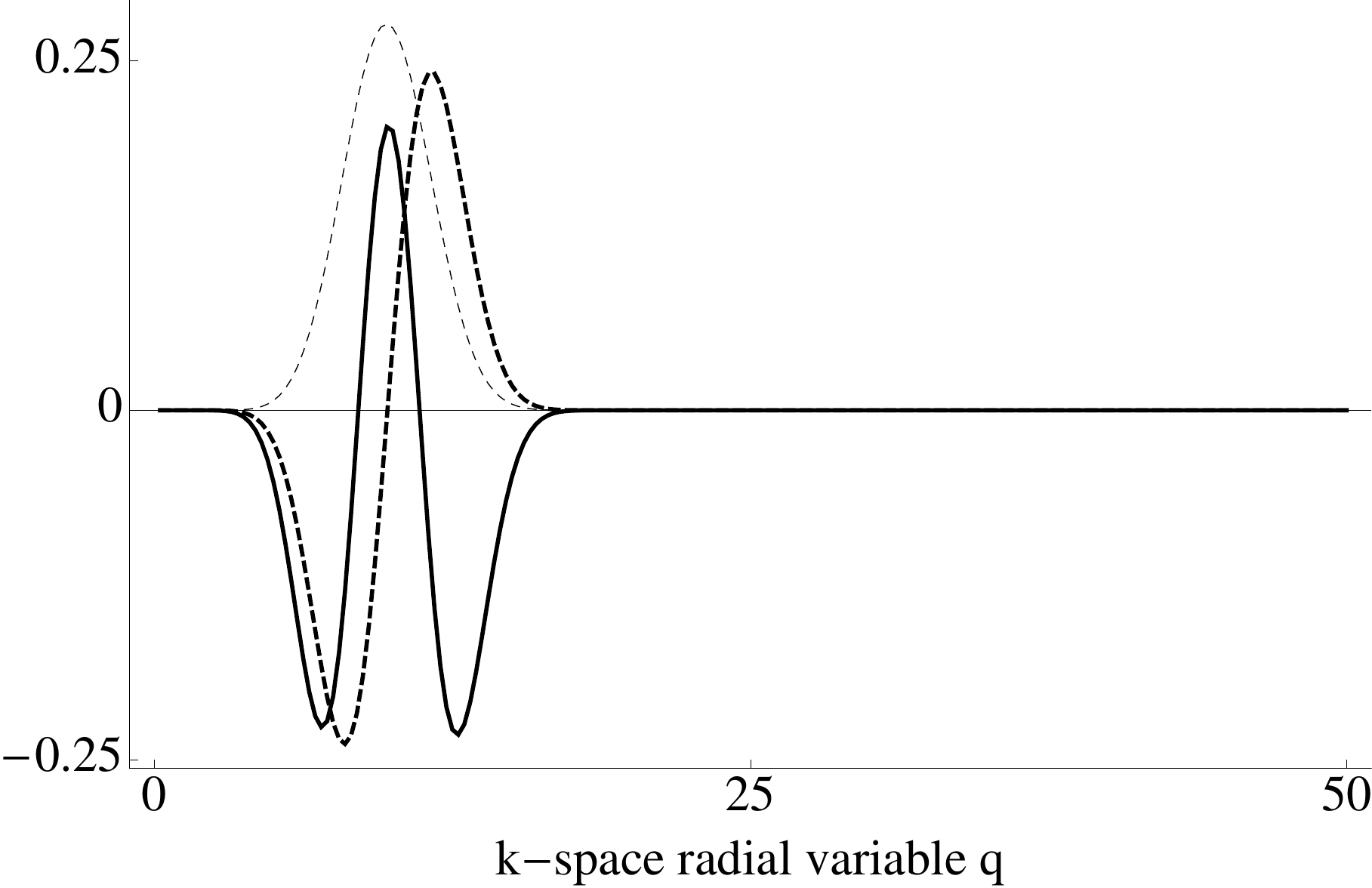}
}
\caption{\label{example0} Amplitudes in the form $\phi_{\ell,p}(q)$ of the first three $p$-modes in figure \ref{fig:eigenmodes}}
\end{figure}

Figure \ref{fig:eigenmodes} shows the result of the singular value decomposition for $b\sigma=0.05$. What is represented is a portion of the matrix $M$ corresponding to a value of the $OAM$ of 10. This choice gives a good example of the radial displacement in $k$-space between Schmidt modes that we observe. Such behaviour will become clear below, where we compare modes from a perfect phase matched case to modes from a negatively detuned phase mismatch. Each horizontal line represents the intensity profile $|\phi_{\ell,p}(q)|^2$ of a radial mode $\phi_{l,p}(q)$ with fixed $\ell$ and a different radial quantum number $p$. The first three modes in figure \ref{fig:eigenmodes} are plotted in figure \ref{example0}.

\begin{figure}[ht]
\resizebox{\columnwidth}{!}{%
\includegraphics{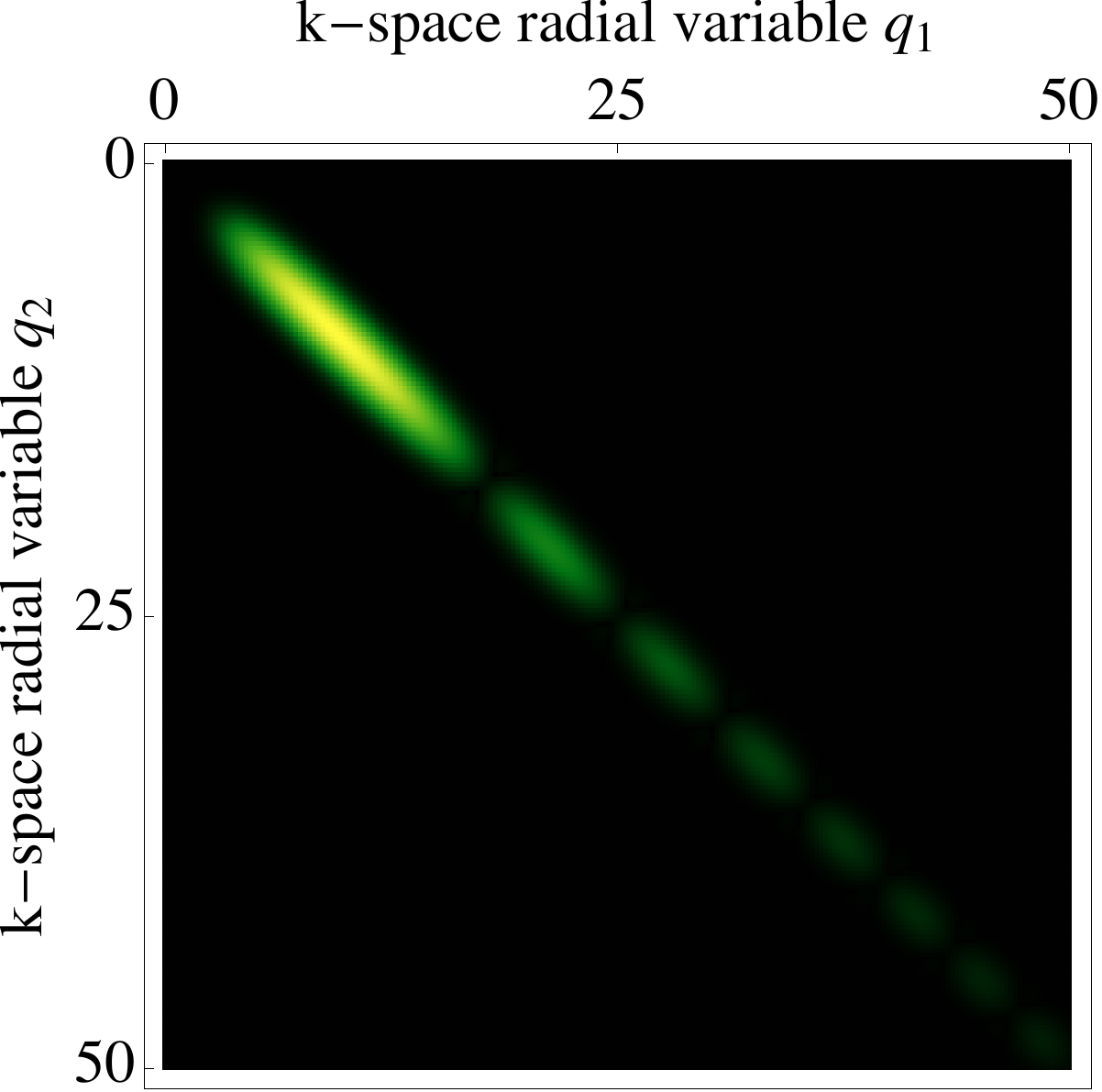}}
\caption{\label{radialfield1} Joint radial probability density of generating a pair of photons with OAM $\ell=10$ and radial momentum $q_1$ and $q_2$. This matrix is the matrix $F$ for $\varphi=0$ used in the examples.}
\end{figure}

\begin{figure}[ht]
\resizebox{\columnwidth}{!}{%
\includegraphics{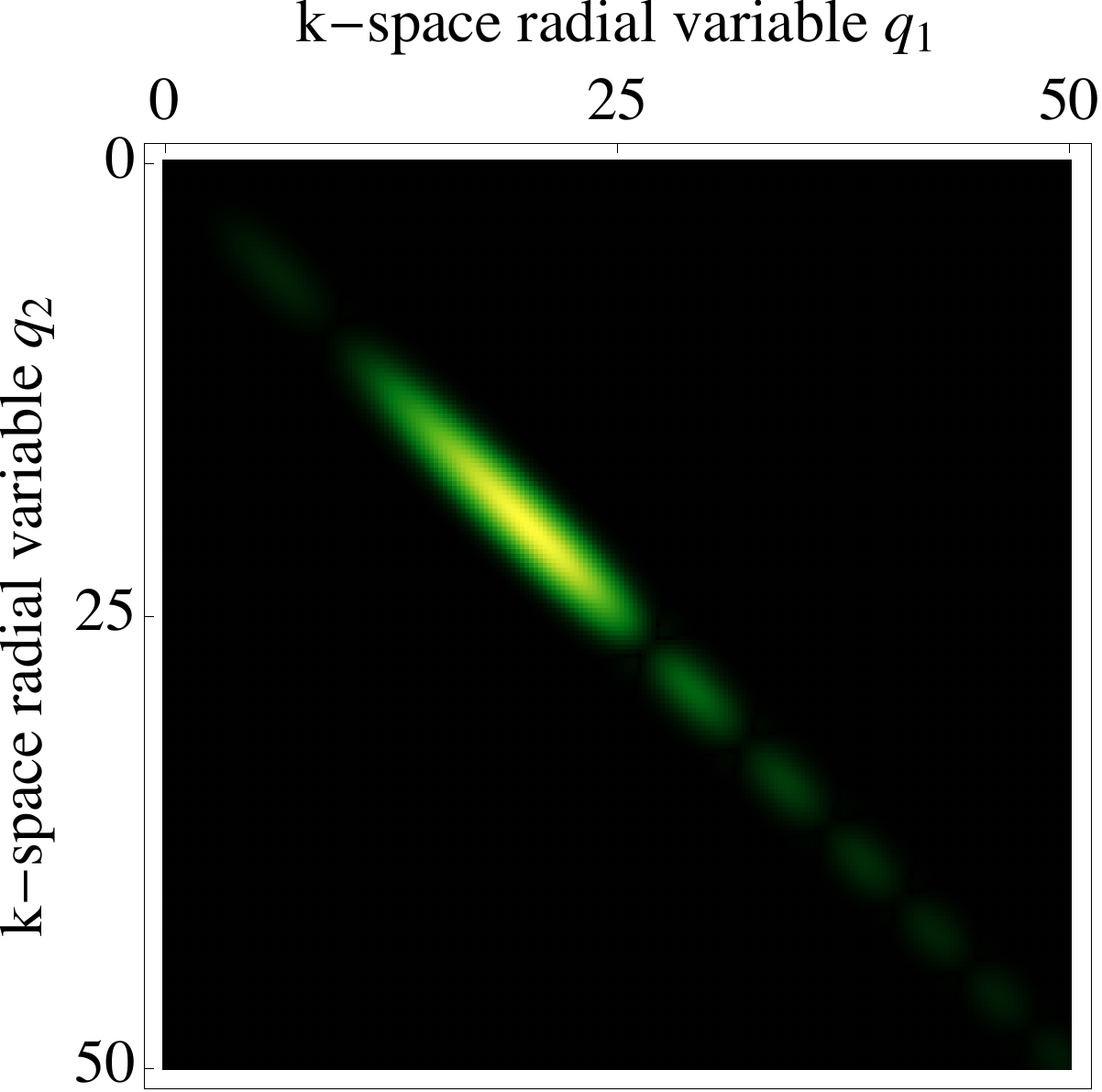}}
\caption{\label{radialfield2} Joint radial probability density of generating a pair of photons with OAM $\ell=10$ and radial momentum $q_1$ and $q_2$. This matrix is the matrix $F$ for $\varphi=-4$ used in the examples.}
\end{figure}

We would like to give an interpretation of the behaviour of the Schmidt modes in figure \ref{fig:eigenmodes}. Notice that each eigenmode resembles a common eigenmode from the family of Hermite-Gauss modes, only centred at specific distances from the origin. There is a one to one correspondence between a family of modes and a region between zeros in the function $\sqrt{P_\ell}F_\ell(q_1,q_2)$. Such function represents the joint amplitude density of the photons having radial momentum $q_1$ and $q_2$, respectively (examples are given in figures \ref{radialfield1} and \ref{radialfield2}). Therefore it's also closely related to the radial sinc profile of the down-converted beam. Such correspondence might be explained by considering the size of the coherence length compared to the size of the nonzero regions in $\sqrt{P_\ell}F_\ell(q_1,q_2)$: if the coherence length is smaller than the thickness of a ring, there can be no strong coherence between photons in different rings, and any wave function in a single ring can be written in terms of a set of fundamental modes of fixed OAM and variable $p$.

We find the organisation of the eigenmodes in groups quite remarkable. Each of the groups looks like a local complete family of HG-like modes, giving the impression of the existence of an additional symmetry and therefore of a quantum number that could be specified to identify each group. This can be explained if the matrix $F^2$ is taken into account and in particular the shape of the lobes that form along the diagonal, as in figure \ref{radialfield1} and \ref{radialfield2}. They can be thought of as potential wells, which have to be filled by eigenmodes. For each of them the eigenmodes build in local families because from the Taylor expansion of the neighbourhood of the bottom of any potential well we can infer that the HG modes form a natural set of local modes for each well. The potential walls between neighbouring wells could be tunneled by the modes, if their size were large enough to reach over to the next well. In practice, the size of each mode is small enough and the families are well confined within individual lobes of the sinc-fucntion if the pump beam is not too focussed.

\subsubsection{Phase mismatch $\varphi\neq0$}
\begin{figure}[ht]
\resizebox{\columnwidth}{!}{%
  \includegraphics{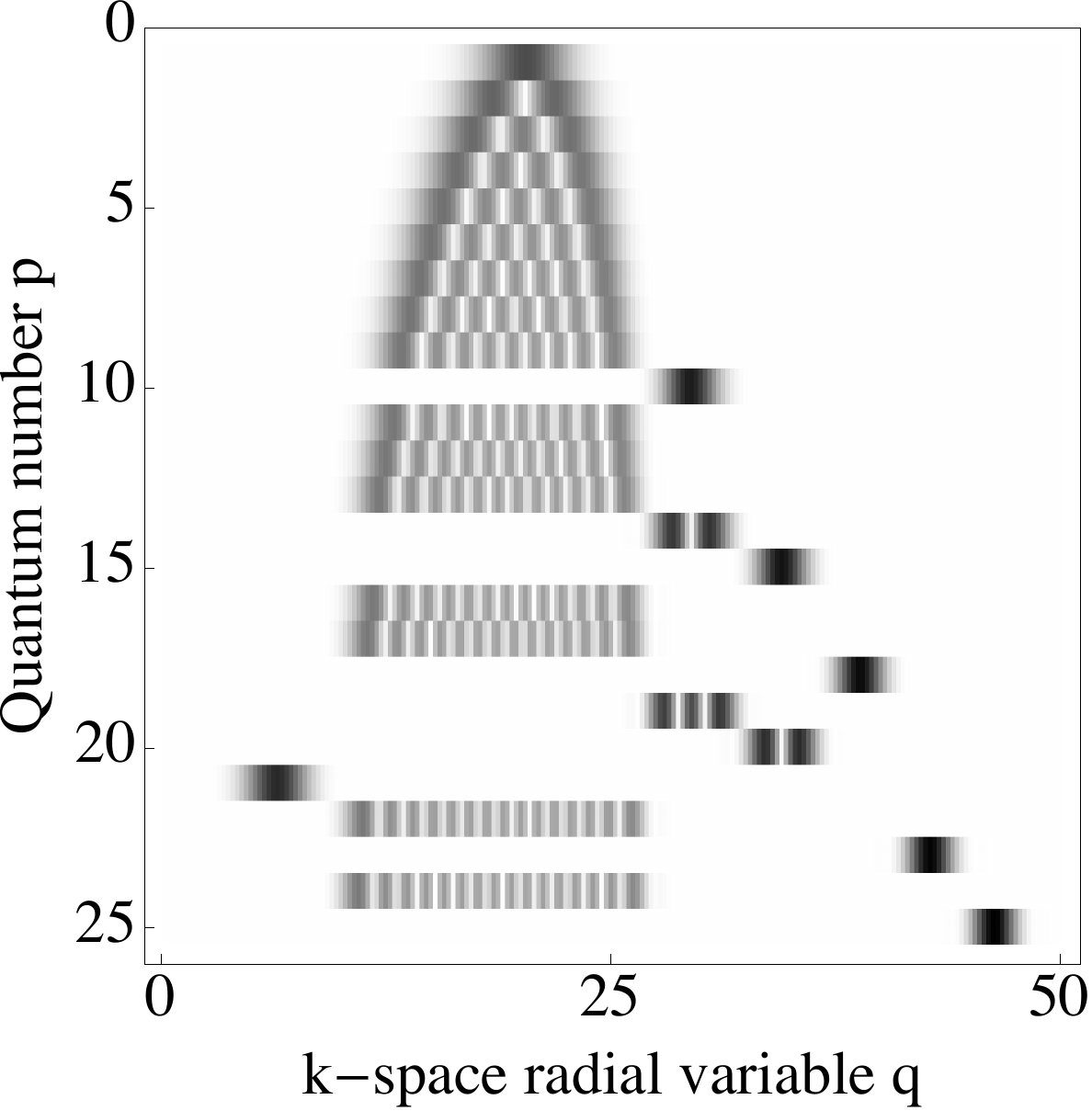}
}
\caption{\label{fig:eigenmodes-4}A part of the matrix $M$ for $\ell=0$ in case of phase mismatch $\varphi=-4$. In this example the radial field that was decomposed was the one in figure \ref{radialfield2}.}
\end{figure}

\begin{figure}[ht]
\resizebox{\columnwidth}{!}{%
  \includegraphics{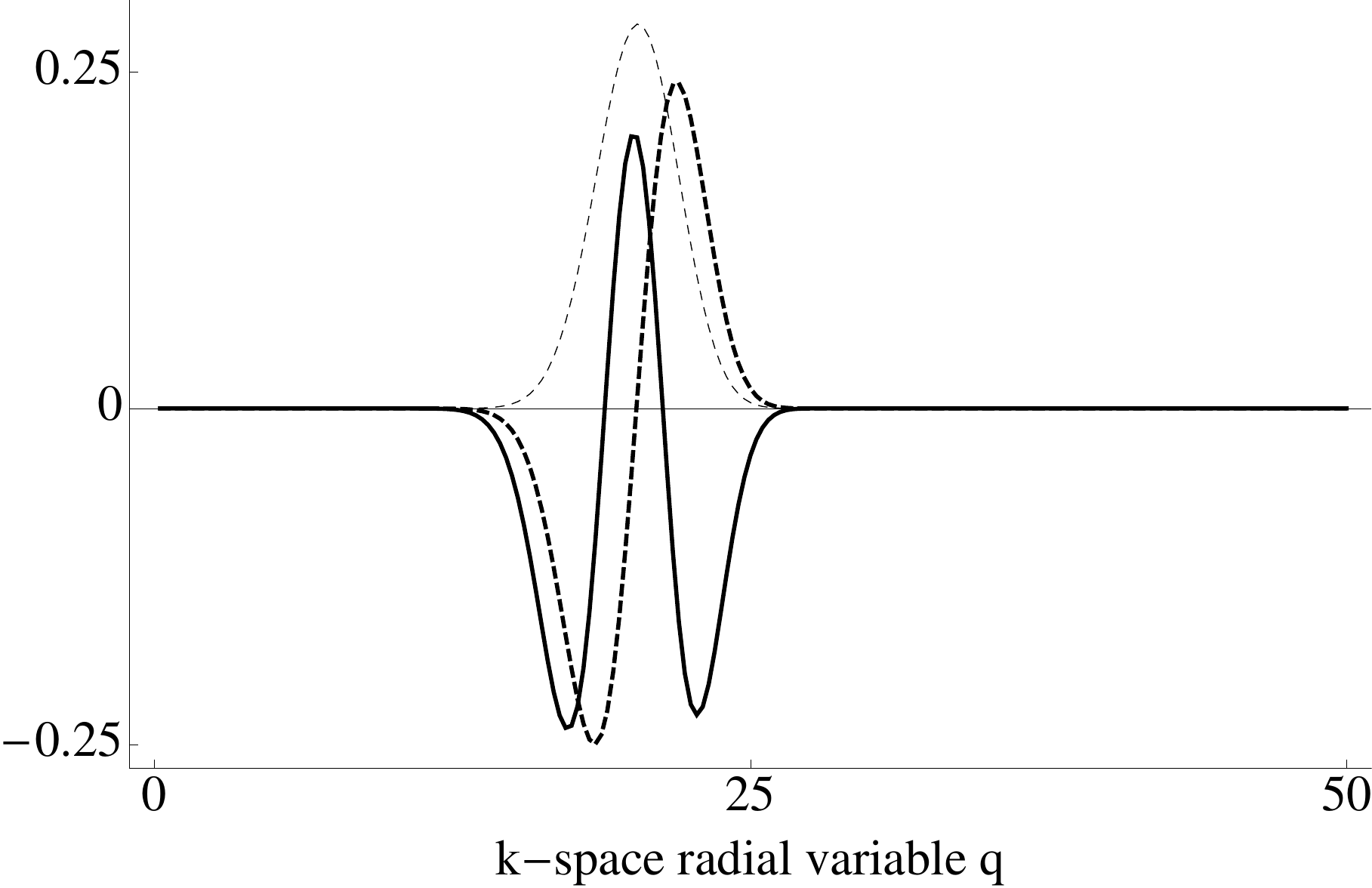}
}
\caption{\label{example-4}   Amplitudes in the form $\phi_{\ell,p}(q)$ of the first three $p$-modes in figure \ref{fig:eigenmodes-4}}
\end{figure}

Figure \ref{fig:eigenmodes-4} shows the calculated radial profile $|\phi_{\ell,p}(q)|^2$ of the eigenmodes at $\ell=10$ for the detuned case $\varphi = - 4$. At $\varphi \ll 0$, where the SPDC ring has opened, even the fundamental $\ell=p=0$ Schmidt mode is ring-shaped and has a wide region of zero intensity around the central axis. As with the zero mismatch example, we also plot the first three modes of figure \ref{fig:eigenmodes-4} in figure \ref{example-4}. Note that although the nonzero phase mismatch, the shape of the first three eigenmodes is remarkably similar to the one in figure \ref{example0}, the difference being just an offset with respect to the origin. 

This happens because in the two cases the radial functions that are being decomposed are remarkably similar. The modes firstly fill the largest ring, until the difference between their combined intensity and the intensity of the main ring is small enough to be less than the intensity of secondary rings. When this happens we see that the subsequent modes fill different rings, and jump discretely between them. This is in accordance with the explanation that we gave at zero phase mismatch: the modes are compact enough to be well contained within individual rings.

\subsection{Heuristic rescaling of Gaussian approximation}
In this section we conclude the argument, anticipated in the previous sections, on the scaling factor $\alpha$. Such scaling factor would allow us to describe more accurately, for $\varphi=0$, the Schmidt number when approximating the sinc phase matching with a Gaussian phase matching. Below, we present two arguments for why we think it is better to replace the function ${\rm sinc}(b^2 q^2)$ by $\exp{(-\alpha^2 b^2 q^2)}$ rather than by $\exp{(-b^2 q^2)}$. There are experimental results that support our claims, for instance see \cite{straupe2011scaling}.

\begin{figure}[htp!]
\resizebox{\columnwidth}{!}{%
  \includegraphics{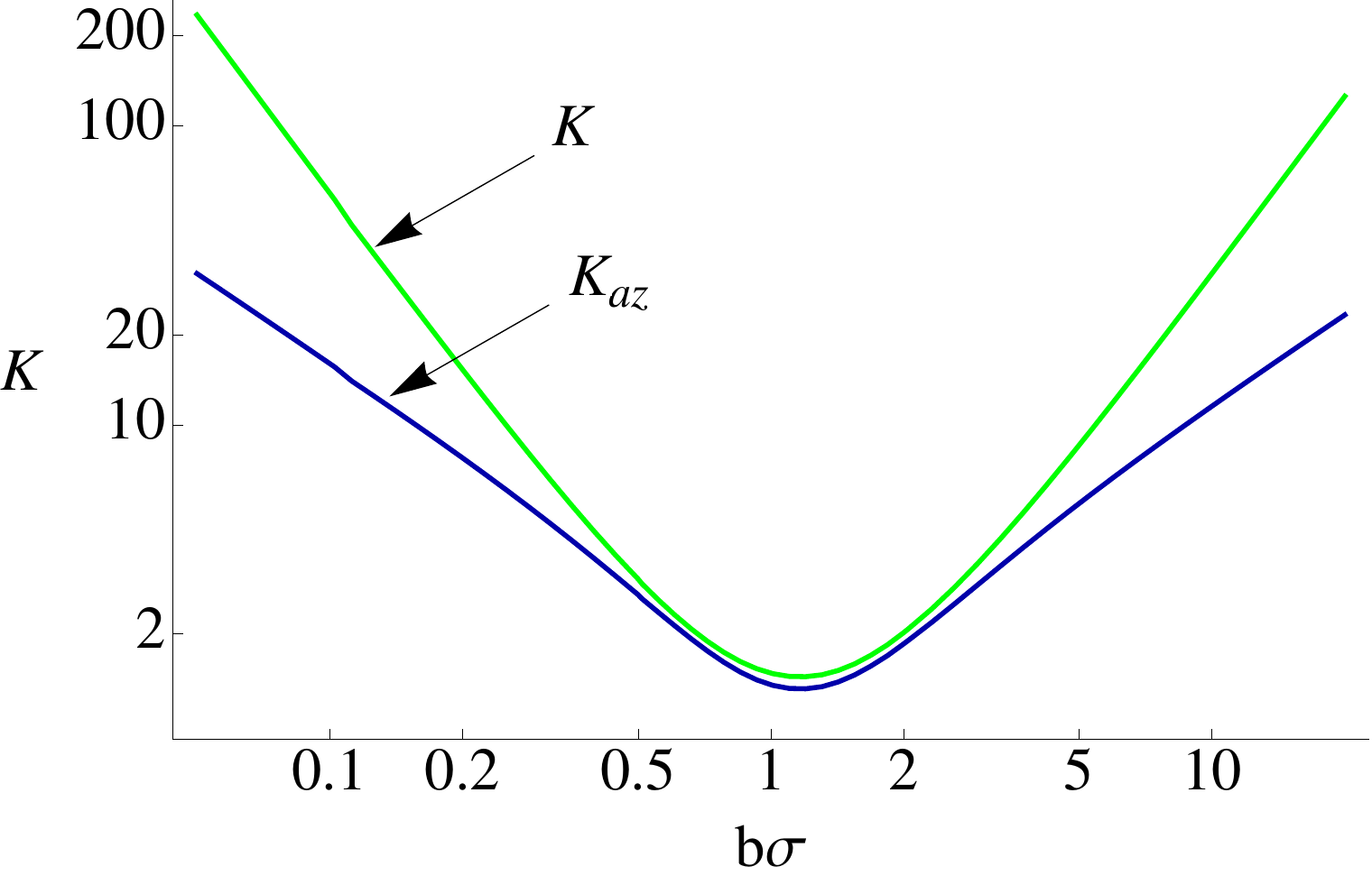}
}
\caption{\label{Kaz_bs}  $K$ and $K_{\mathrm{az}}$ as functions of $b\sigma$ for $\varphi=0$.}
\end{figure}

\begin{figure}[htp!]
\resizebox{\columnwidth}{!}{%
  \includegraphics{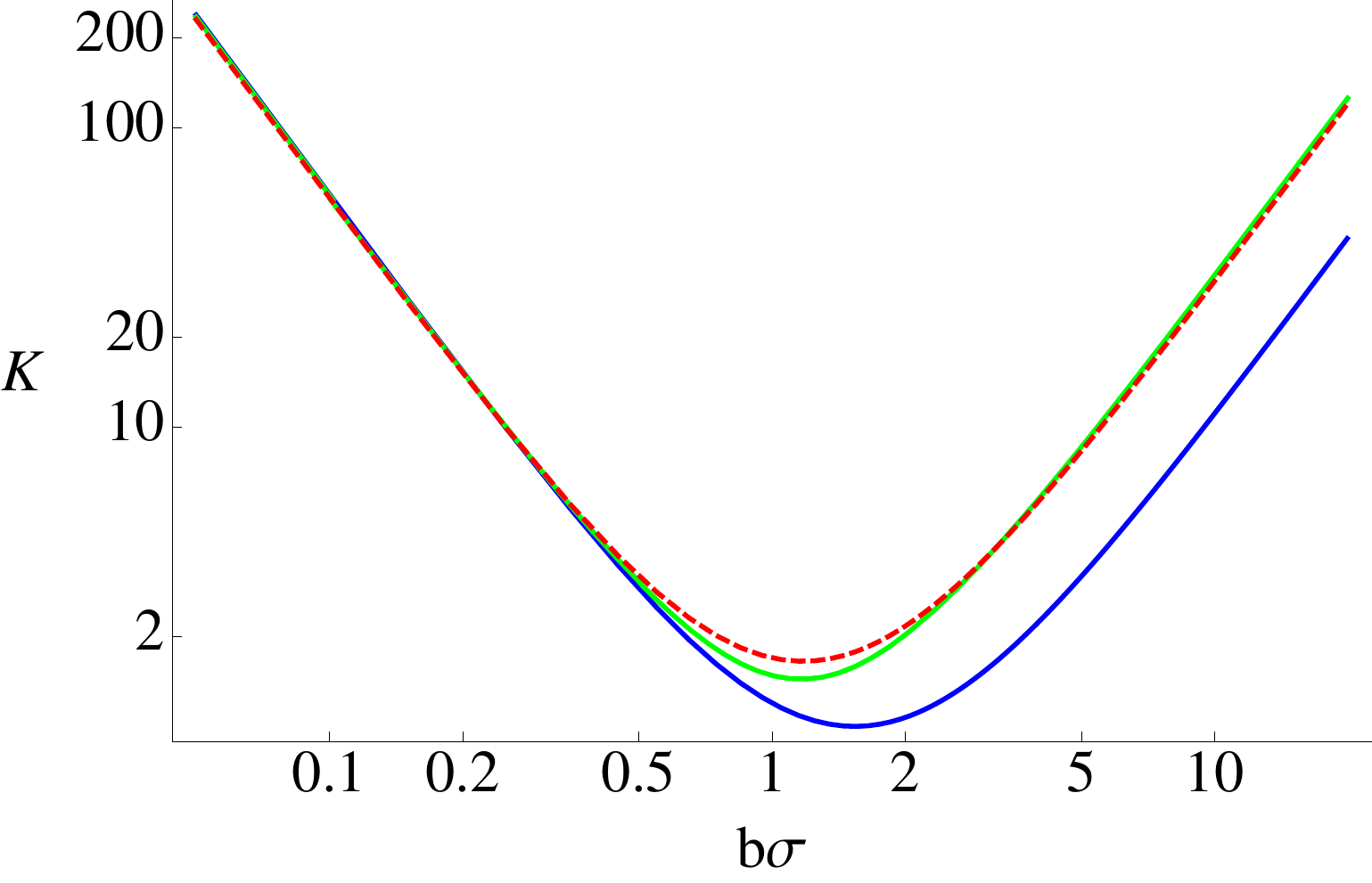}
}
\caption{\label{Kaz_bs_mod}  $K$ in green and the two different modifications of the law \eqref{K}. The dashed red line is the $(\alpha,\beta)$-modification, the solid blue line is the $\alpha$-only modification. The fits obviously match for $b\sigma\lesssim 1$.}
\end{figure}

Figure \ref{Kaz_bs} depicts the numerically-calculated Schmidt number $K$ and its azimuthal counterpart $K_{\mathrm{az}}$ as functions of $b \sigma$ for the exact sinc phase matching. In order to numerically calculate these curves we had to abandon momentarily the faster approach used to find the Schmidt modes, as $b\sigma$ in figure \ref{Kaz_bs} is not restricted to small values. The Gaussian approximation of the phase matching function results in expressions that are symmetric under the exchange $b\sigma\leftrightarrow\frac{1}{b\sigma}$.  The same argument cannot be applied to the sinc phase matching, as it's possible to see from figure \ref{Kaz_bs}: the graph is symmetric with respect to a point between 1 and 2, meaning that there could be an extra factor $\alpha$ such that the symmetry now reads $b\sigma\alpha\leftrightarrow\frac{1}{b\sigma\alpha}$, representing a reflection around the point $1/\alpha$. If it were so, the factor $\alpha$ should have a value around $\alpha\simeq0.85$. However, the introduction of $\alpha$ alone is not enough to obtain the correct $K$ values. There are two ways to proceed, either to choose a simple modification, like $b\sigma\rightarrow b\sigma\alpha$, or to add also a second multiplicative factor $\beta$. We find that the relation

\begin{align}
K=\frac{\beta}{4}\left(\frac{1}{b\sigma\alpha}+b\sigma\alpha\right)^2
\label{alphabeta}
\end{align}

fits extremely well the values in figure \ref{Kaz_bs} for $\alpha=0.85$ and $\beta=1.65$. However, if one is only interested in the regime $b\sigma \lesssim 1$, which experimentally is the most signiÞcative, there is no need to introduce $\beta$ and the relation $K \approx 1/(2 b\sigma \alpha')^2$ works well for $\alpha' \equiv \alpha/\sqrt{\beta} \approx 0.65$, as it is possible to see in figure \ref{Kaz_bs_mod}. The $\alpha$-modification is a consistent choice for replacing the sinc phase matching with a Gaussian phase matching, as it is implemented by modifying the width of the Gaussian and not by modifying the relation between $K$ and the product $b\sigma$ ad hoc, as in \eqref{alphabeta}. 

The second argument relays on a common criterium to choose a sensible value for $\alpha$ without recurring to ad hoc arguments. The common criterium is equal width at $1/e$ from the peak intensity \cite{Joobeur1996}. This criterium yields again a value of $\alpha\simeq0.65$.

%%%%%%%%
%Conclusions
%

\section{Conclusions}
We developed a fast numerical approach to perform a numerical Schmidt decomposition without approximating the phase matching function, and for any phase mismatch. Thanks to the numerical approach we investigated the limitations of the Gaussian approximation and we proposed a scaling $b\rightarrow b\alpha$, for $\alpha\simeq0.65$, that allows the Gaussian approximation with no phase mismatch to reproduce the results of the more complete sinc phase matching. We also successfully verified the etendue argument by calculating $K$ and $K_{az}$ for different values of the phase mismatch.

The Schmidt modes that we found exhibit numerous features, the most remarkable being the grouping in local families of HG-like modes and the fact that they don't span more than one lobe on the diagonal of the matrices of joint radial probability density. We gave an interpretation of these features by using the meaning of coherence length and the way of naturally assigning eigenmodes to potential wells. In fact, the other sets of HG-like radial modes are simply shifted by different amounts into other rings of the sinc-type SPDC emission. These other sets have similar HG-like radial structures but lower modal weights. The rather strict separation in discrete sets suggests that we might want to use the set number as third quantum number, i.e. to split the radial quantum number $p$ in two quantum numbers instead.

\section*{Appendix}
Here we report more in detail the function that we use to fill the matrix $F$. We start with the following two-photon wave function

\begin{align}
\Psi(\mathbf{q}_1,\mathbf{q}_2)=\mathcal Ne^{-\frac{w_0^2}{4}|\mathbf{q}_1+\mathbf{q}_2|^2}\mathrm{sinc}\left(\Phi+\frac{L}{4k_p}|\mathbf{q}_1-\mathbf{q}_2|^2\right)
\end{align}
which can be written in cylindrical coordinates:
\begin{align}
\hspace{-1.5cm}&\Psi(q_1,\varphi_1,q_2,\varphi_2)=\nonumber\\&=\mathcal N\exp\left(-\frac{w_0^2}{4}(q_1^2+q_2^2)-\underbrace{\frac{w_0^2}{2}q_1q_2}_\alpha\cos(\varphi_1-\varphi_2)\right)\nonumber\\ &\times\mathrm{sinc}\left(\underbrace{\Phi+\frac{L}{4k_p}(q_1^2+q_2^2)}_\beta-\underbrace{\frac{L}{2k_p}q_1q_2}_\gamma\cos(\varphi_1-\varphi_2)\right)\\
&=\mathcal N'\exp(-\alpha\cos(\varphi_1-\varphi_2))\mathrm{sinc}(\beta-\gamma\cos(\varphi_1-\varphi_2))
\end{align}
where $\mathcal N'=\mathcal N\exp(-\frac{w_0^2}{4}(q_1^2+q_2^2))$

Similarly to what was done for the exp phase matching, we begin the evaluation of the Fourier transform with $\exp(-i\ell(\varphi_1-\varphi_2))$ as the Fourier kernel, which will give the eigenfunctions of a state of OAM $\pm \ell$.

\begin{align}
\sqrt{P_\ell}F(q_1,q_2)&=2\pi\mathcal N'\underbrace{\int_0^{2\pi}e^{-\alpha\cos\varphi}\mathrm{sinc}(\beta-\gamma\cos\varphi)e^{-i\ell\varphi}\,d\varphi}_{H_\ell}
\end{align}

We have

\begin{align} 
H_\ell&=\int_0^{2\pi}e^{-\alpha\cos(\varphi)}\mathrm{sinc}(\beta-\gamma\cos(\varphi))e^{-i\ell\varphi}\,d\varphi\\
&=\frac{1}{L}\int_{-L/2}^{L/2}e^{\frac{2i\beta}{L}t}\int_0^{2\pi}e^{-(\alpha+\frac{2i\gamma}{L}t)\cos(\varphi)-i\ell\varphi}d\varphi dt\\
&=\frac{2\pi}{L}\int_{-L/2}^{L/2}e^{\frac{2i\beta}{L}t} I_{|\ell|}(-\alpha-\frac{2i\gamma}{L}t)\,dt\\
&=\pi\int_{-1}^{1}e^{i\beta t}I_{|\ell|}(-\alpha-i\gamma t)\,dt
\label{H}
\end{align}

This last form of the integral can be approximated if the product $b\sigma$ can be considered sufficiently small. The amount of precision sought restricts the maximum value of $L_R$, but at $L_R<0.1$ (which corresponds to $b\sigma\lesssim0.2$) the maximum error is less than 1\%. The approximating function of $e^{-\frac{w_0^2}{4}(q_1^2+q_2^2)}H_\ell$ is found to be
\begin{align}
f(q_1,q_2)=2\pi e^{-(q_1^2+q_2^2)}I_\ell(2 q_1 q_2)\, \mathrm{sinc}\left[\sqrt{2L_R} \left(q_1^2+q_2^2\right)+\Phi \right]
\end{align}
A matrix $F$ is filled with entries $(i,j)=s\sqrt{ij}f(is,js)$, where $s$ is the step, until the size $N$ of the matrix is reached. Then $F^2=FF^\dag$ is used in the singular value decomposition. The resulting eigenvalues are $\lambda_{\ell,p}$ (at fixed $\ell$); the resulting eigenmode are $\phi_{\ell,p}(q) = \sqrt{q} u_{\ell,p}(q)$.

At sufficiently large x, the modified Bessel function can be expanded as $I_\ell(x) \approx e^x/\sqrt{2\pi x}$. By combining this expansion with the mentioned multiplication by $\sqrt{q_1 q_2}$, we find that the matrix $F$ has a similar appearance as the original two-photon field of Eq.(4), but with the roles of $(q_1-q_2)$ and $(q_1+q_2)$ interchanged and a removal of the vector character. After Taylor expansion of the sinc-function, which works best in the considered limit $L_R << 1$ for $\Phi < 0$, the resulting bi-exponential function yields a set of Hermite-Gaussian radial Schmidt modes centered around a displaced maximum.

\section*{Acknowledgements}
This work was supported by the UK EPSRC.
We acknowledge the financial support of the Future and Emerging
Technologies (FET) program within the Seventh Framework Programme
for Research of the European Commission, under the FET Open grant
agreement HIDEAS number FP7-ICT-221906.
This research was supported by the DARPA InPho program through the US Army Research Office award W911NF-10-1-0395. 

%%%%%%%%
%Bibliography
%

\bibliographystyle{epj}
\bibliography{Schmidtmodes}

\begin{thebibliography}{25}

\bibitem{Neilsen}
M.A. Nielsen, I.L. Chuang, \emph{Quantum computation and quantum information}
  (Cambridge University Press, Cambridge, 2000)

\bibitem{Sonja:2002}
S.~Franke-Arnold, S.M. Barnett, M.J. Padgett, L.~Allen, Phys. Rev. A
  \textbf{65}, 033823 (2002)

\bibitem{Torres:2003}
J.P. Torres, A.~Alexandrescu, L.~Torner, Phys. Rev. A \textbf{68}, 050301
  (2003)

\bibitem{Molina-Terriza:2008}
G.~Molina-Terriza, J.P. Torres, L.~Torner, Nat. Phys. \textbf{3}, 305 (2008)

\bibitem{Franke-Arnold:2008}
S.~Franke-Arnold, L.~Allen, M.J. Padgett, Laser Photonics Rev. \textbf{2}, 299
  (2008)

\bibitem{Miatto2012}
F.M. Miatto, T.~Brougham, A.M. Yao, Eur. Phys. J. D \textbf{66}, 1 (2012)

\bibitem{Walmsley2000}
C.K. Law, I.A. Walmsley, J.H. Eberly, Phys. Rev. Lett. \textbf{84}, 5304 (2000)

\bibitem{Straupe2011}
S.S. Straupe, D.P. Ivanov, A.A. Kalinkin, I.B. Bobrov, S.P. Kulik, Phys. Rev. A
  \textbf{83}, 060302 (2011)

\bibitem{Mandelwolf}
L.~Mandel, E.~Wolf, \emph{Optical Coherence and Quantum Optics} (Cambridge
  Univ. Press, 1995)

\bibitem{Law2004}
C.~Law, J.~Eberly, Physical Review Letters \textbf{92}, 127903 (2004)

\bibitem{Alison2010b}
A.M. Yao, New Journal of Physics \textbf{13}(5), 053048 (2011)

\bibitem{Monken1998}
C.H. Monken, P.H.S. Ribeiro, S.~P\'adua, Phys. Rev. A \textbf{57}, 3123 (1998)

\bibitem{Exter2006}
M.P. van Exter, A.~Aiello, S.S.R. Oemrawsingh, G.~Nienhuis, J.P. Woerdman,
  Phys. Rev. A \textbf{74}, 012309 (2006)

\bibitem{Walmsley2003}
A.B. U'Ren, K.~Banaszek, I.A. Walmsley, Quantum Inf. Comput. \textbf{3}, 480
  (2003)

\bibitem{Walmsley2010}
L.E. Vicent, A.B. U'Ren, R.~Rangarajan, C.I. Osorio, J.P. Torres, L.~Zhang,
  I.A. Walmsley, New Journal of Physics \textbf{12}, 093027 (2010)

\bibitem{korsch}
H.J. Korsch, A.~Klumpp, D.~Witthaut, quant-ph/0608216v1  (2006)

\bibitem{dattoli}
G.~Dattoli, C.~Chiccoli, S.~Lorenzutta, G.~Maino, M.~Richetta, A.~Torre, J.
  Sci. Comp. \textbf{8}, 69  (1993)

\bibitem{Pires2009c}
H.~{Di Lorenzo Pires}, C.H. Monken, M.P. van Exter, Physical Review A
  \textbf{80}, 22307 (2009)

\bibitem{Pires2009b}
H.~{Di Lorenzo Pires}, M.~van Exter, Physical Review A \textbf{79}(4), 1
  (2009), ISSN 1050-2947

\bibitem{HOM}
C.~Hong, Z.~Ou, L.~Mandel, Physical Review Letters \textbf{59}, 2044 (1987)

\bibitem{Zambrini2006}
R.~Zambrini, S.M. Barnett, Phys. Rev. Lett. \textbf{96}, 113901 (2006)

\bibitem{Pires2010}
H.D.L. Pires, J.~Woudenberg, M.P. van Exter, Optics letters \textbf{35}, 889
  (2010)

\bibitem{Miatto2012b}
F.M. Miatto, D.~Giovannini, J.~Romero, S.~Franke-Arnold, S.M. Barnett, M.J.
  Padgett, Eur. Phys. J. D \textbf{66}, 178 (2012)

\bibitem{straupe2011scaling}
S.S. Straupe, D.P. Ivanov, A.A. Kalinkin, I.B. Bobrov, S.P. Kulik,
  quant-ph/1112.3806v1  (2011)

\bibitem{Joobeur1996}
A.~Joobeur, B.~Saleh, T.~Larchuk, M.~Teich, Physical Review A \textbf{53}, 4360
  (1996)

\end{thebibliography}

\end{document}